\documentclass[11pt]{article}
\usepackage{moriond,epsfig}

\usepackage{latexsym,amssymb,psfrag}
\usepackage{graphicx}
\usepackage{pstricks}
\usepackage{amsmath}
\usepackage{subfigure}

\def\be{\begin{equation}}
\def\ee{\end{equation}}
\def\bea{\begin{eqnarray}}
\def\eea{\end{eqnarray}}

\begin{document}
\vspace*{4cm}
\title{Dark Matter searches and Colliders 
%\footnote{Talk given at the XXXXth Rencontres de Moriond session devoted to
%ELECTROWEAK INTERACTIONS AND UNIFIED THEORIES, March 5-12, 2005, La Thuile
%(ITALY).}
}

\author{E.  Nezri }

\address{Service de Physique Th\'eorique, Universit\'e Libre de Bruxelles B-1050 Brussels, Belgium }

\maketitle\abstracts{We study the potential of neutralino dark matter searches
  and  SUSY particle production at colliders in the framework of the Minimal Supersymmetric Standard Model. We
  show the effects of a non universal gluino mass parameter. The
  future experiments will be a stringent probe of the low energy MSSM and
  neutralino dark matter scenario. 
The deeper informations on both supersymmetry and astrophysics hypothesis will
  be obtained by correlation of the different signals or absence of signal.}

\section{Introduction}

Several astrophysics indications point that matter in the universe is
 dominated by an unidentified and undiscovered dark matter (see
 {\it e.g} \cite{Bertone:2004pz,Jungman:1995df,carlosreview} for a
 review).  The WMAP results lead to a flat $\Lambda CDM$ universe with
 $\Omega_{CDM} h^2=0.1126^{\ +0.0161}_{\ -0.0181}$.
An interesting possibility for the dark matter enigma is a bath of
 Weakly--Interacting Massive Particle (WIMP).

In high energy physics, the most ppopular extension of the Standard Model is achieved through the
 supersymmetry (SUSY) suggested by theoretical aspects and string theory. The
  Minimal Supersymmetric Standard Model (MSSM) \cite{Fayet:1976cr,Haber:1984rc,Barbieri:1987xf}, predicts the existence of
 several new particles, the superpartners of known particles. Those particles
 should be discovered in future collider experiments. Furthermore the  $L$ightest $S$upersymmetric $P$article
($LSP$) is for most of the MSSM parameters a stable (assuming
 R-parity conservation), massive, neutral and weakly
 interacting particle : the lightest neutralino being so an interesting and
 well motivated candidate for dark matter.

The neutralinos ($\chi^0_1\equiv \chi,\chi^0_2,\chi^0_3,\chi^0_4,$) are the masses eigenstates coming from mixing of neutral gauge
and (Brout-Englert-)Higgs boson superpartners:  $\tilde B$, $\tilde {W}^3$,
$\tilde{H}^0_1$, $\tilde{H}^0_2$ respectively the B--ino, W--ino and up, down--Higgsinos fields.  In the
($\tilde{B},\tilde{W}^3,\tilde{H}^0_1,\tilde{H}^0_2$) basis, the neutralino mass matrix is

\begin{equation}
\arraycolsep=0.01in
{\cal M}_N=\left( \begin{array}{cccc}
M_1 & 0 & -m_Z\cos \beta \sin \theta_W^{} & m_Z\sin \beta \sin \theta_W^{}
\\
0 & M_2 & m_Z\cos \beta \cos \theta_W^{} & -m_Z\sin \beta \cos \theta_W^{}
\\
-m_Z\cos \beta \sin \theta_W^{} & m_Z\cos \beta \cos \theta_W^{} & 0 & -\mu
\\
m_Z\sin \beta \sin \theta_W^{} & -m_Z\sin \beta \cos \theta_W^{} & -\mu & 0
\end{array} \right)\;.
\label{eq:matchi}
\end{equation}

\noindent
 $M_1$, $M_2$
and $\mu$ are the bino, wino, and Higgs--Higgsino mass parameters respectively
after Renormalization Group Equation (RGE) running from GUT scale down to
electroweak symmetry breaking (EWSB) scale.
Tan$\beta$ is the ratio of the $vev$ of the two Higgs doublet fields.
This matrix can be diagonalized by a single unitary matrix $z$ 
such that we can express the LSP $\chi$ ({\it the neutralino} in the following) as

\begin{equation}
\chi = z_{1 1} \tilde B+  z_{1 2}\tilde W  
+ z_{1 3}\tilde H_1  +z_{1 4} \tilde H_2.
\end{equation}

\noindent 
This combination determines the nature, the couplings and the phenomenology of
 the neutralino.

In this talk, we will consider neutralino dark matter searches and SUSY
particles production in future colliders. 
Related works in a variety of framework models treating relic density aspect, present accelerator constraints and/or
DM searches and/or SUSY searches in future colliders can be found in {\it
  e.g.} references  \cite{ellissug1} - \cite{Baer:2005zc}.

\section{Dark matter searches}

\subsection{Dark matter distribution}

The dark matter distribution is an important point for detections.
 If there is an agreement concerning the behavior at
large radii, the shape of the innermost region of the
galaxy is quite uncertain if we consider the discrepancies between simulation
 results of various groups. Further, the observation of systems like low 
surface
brightness seem to favor flat cores. On top of that the small radius region 
behavior can differ strongly depending on physics assumptions considered
like effects of the baryons, supermassive black hole induced spike,
 dark matter particle scaterings by stars.
On the contrary, the local density $\rho_0$ is more under control and should 
be in the range $0.2-0.8\ {\rm GeV}.{\rm cm}^{-3}$.
Finally possible inhomogeneities and substructures could be present giving a
possible clumpyness of the halo.

\subsection{Direct detection}

The presence of dark matter particles can be signed by the collision with the nuclei of a detector. 
The astrophysical dependence is weak and comes from the knowledge of local
dark matter density $\rho_0$.  From the particle physics point of view, depending on the spin of the target
nuclei, the detection is  driven by the spin dependent or scalar neutralino
proton(nucleon) elastic cross section {\it i.e} processes in $\sigma_{\chi-q}$.

Current experiments like  EDELWEISS and CDMS are sensible to WIMP--proton cross
section $\lesssim 10^{-6}$ pb which
is slightly not enough to probe SUSY model including RGE and  radiative EWSB if one
requires experimental constraints and near WMAP relic density. 
The next step of experiments ({\it e.g } EDELWEISS II and CDMS II) will lead
to a minimum of the valley sensitivity around $10^{-8}$ pb for $m_{\chi}$ of
order 100 GeV. Though challenging from the experimental point of view, a
ton-size detector (ZEPLIN, SuperCDMS) should be able to reach  $\sigma^{scal}_{\chi-p} \lesssim
10^{-10}$ pb which would be a conclusive tool to probe WIMP dark
matter models especially for the MSSM neutralino scenario.

\subsection{Gamma Indirect detection}

Dark matter can annihilate in the halo, especially in the
Galactic Center where the DM density is high. This can lead to important gamma
fluxes and promising signals despite of the strong uncertainty coming from 
 the unknown distribution in the innermost region.
The particle physics dependence comes from neutralino annihilation cross
section. 
There are several existing signals (INTEGRAL, EGRET, CANGAROO, HESS) from the Galactic Center region extending on very 
different energy ranges.  Though possible explanation in term of neutralino (see {\it e.g}
\cite{deBoer:2004ab}, \cite{Aharonian:2004wa}) are
possible for each signal (except {\it maybe} for INTEGRAL, see \cite{Boehm:2003bt} for the
light dark matter proposition), those measurements ares not
compatible with each other and can not be explained by a single scenario. 
Nevertheless, the Egret signal  ($\sim 4\times 10^{-8}\ \gamma{\rm cm^{-2}.s^{-1}}$) can represent an upper
bound. We will consider the HESS and GLAST experiment sensitivities
(respectively $\sim 10^{-12-13}\ \gamma\ {\rm
  cm^{-2}.s^{-1}}$ with 60 GeV threshold and $\sim 10^{-10}\ \gamma\ {\rm
  cm^{-2}.s^{-1}}$ with 1 GeV threshold) as a
probing test of SUSY models.

\subsection{Neutrino Indirect detection}

Dark matter particles of the halo can also be trapped in the Sun by successive elastic diffusion on its nuclei (Hydrogen). This lead to a
captured population which can annihilate. The annihilation products then decay
producing neutrino fluxes which can be detected by a neutrino telescopes
signing the presence of dark matter in the Sun.

The local dark matter density is the weak astrophysical dependence for this
possible detection and it is shared with direct detection. 
Concerning particle physics dependence, the main aspect is the capture rate
driven by $\sigma_{\chi -q}$ and neutralino annihilation cross section.

Present experiments like MACRO, BAKSAN, SUPER
K and AMANDA  give limits on possible fluxes around $10^4\ \mu\
{\rm km^{-2}.yr^{-1}}$. Future neutrino telescopes like ANTARES  or a ${\rm
  km}^3$ size like ICECUBE will be able to probe respectively around
$10^3$ and $10^2\ \mu\ {\rm km^{-2}.yr^{-1}}$.

\subsection{Positron and Antiproton Indirect detection}

Neutralino annihilations in the halo can also give rise to measurable positron
and antiproton
fluxes.  Being charged particles, they interact during their
propagation such that the directional information is lost.  The variability at the production level
due the density profile  is quite weak, indeed the understanding of
the propagation and the resolution of diffusion equation is the most relevant 
issue. The particle physics dependence enters in the source term by the 
annihilation cross section. 

The HEAT experiment has measured an excess of positron (peaked
around 10 GeV)  which can be accommodated by neutralino 
but requires a boost factor \cite{Baltz:2001ir,hoopersilk}. We will probe SUSY
models with regard to the future 
experiments AMS-02 and PAMELA. Considering the positron spectra being peaked around
$M_{\chi}/2$, a benchmark condition can be \cite{Feng,Baer:2004qq}:  $
\frac{\phi^{e^+}_{\chi}}{\phi^{e^+}_{Bckgd}}|_{M_{\chi}/2} \sim 0.01 $. 
The antiproton flux is measured by experiments like BESS and CAPRICE and is
accommodated by astrophysical processes. This flux is peaked at 1.76 GeV around $2\times 10^{-6}\
\bar{p}\ {\rm cm^{-2}s^{-1}sr^{-1}}$. We will 
show as benchmark the region where $\phi_{\bar{p}}(R_0,T)> 2\times 10^{-7} \
\bar{p}\ {\rm cm^{-2}s^{-1}sr^{-1}}$.

\section{Collider searches}

\subsection{LHC}

%genevieve {\bf [to be completed/changed by yann and jean-bapt ]}

As a probe of supersymmetric parameter space, we consider sparticles production
in which strong interaction is relevant {\it i.e} squarks and gluinos
production: $parton-parton\xrightarrow{g} \tilde{q} \tilde{q},\  \tilde{g}
\tilde{q},\  \tilde{g} \tilde{g} $. The decays of squarks and gluinos lead to multi-jets + isolated leptons +
missing $E_T$ signals. We consider the exclusion limits of
reference\cite{Charles:2001ka} which establishes that squarks and gluinos could be
detected up to $m_{\tilde{q} -\tilde{q}} \sim 2-2.5$ TeV.

\subsection{LC}

%genevieve  {\bf [to be completed/changed by yann and jean-bapt ]}

We also probe supersymmetric models with regard to a possible Linear
Collider with an energy of 1 TeV,  $L= 500 fb^{-1}/yr$ and require 50 events.\\ We consider the following processes: $e^+e^-\xrightarrow{}\tilde{l}\tilde{l},\ \tilde{\chi^+}\tilde{\chi^-},\
\tilde{\chi}\tilde{\chi^0_2},\ HA$.

\section{Prospection}

We will now show the regions in supersymmetric parameter space which are
accessible for typical experiments (red and green) of the different kinds of 
detection. The areas excluded by the experimental constraints(Higgs and
chargino mass, $ b \rightarrow s \gamma$, muon anomalous magnetic moment, $
B_s \rightarrow \mu^+\mu^-$) are also shown (grey). Considering possible
 alternatives in cosmology to the standard thermal DM scenario and
 particle physics uncertainties in the calculation of the MSSSM spectrum and
 the relic density, we show
 the conservative range $0.03<\Omega_{\chi}{\rm h}^2<0.3$
 (yellow with external black lines), as well as the WMAP range (internal
black lines of the yellow areas).
Our starting parameter space is the Universal mSugra/CMSSM plane, where one
assumes  a unified gaugino mass at GUT scale ($m_{1/2}$) and a unified scalar
mass at GUT scale ($m_{0}$). As a benchmark, we choose
$A_0=0,\tan{\beta}=35,\mu>0$. In this proceeding we will only illustrate the 
effect 
of non universal gluino mass term $M_3|_{GUT}$. The effects of non universal wino
mass $M_2|_{GUT}$, up-type Higgs mass $MH_u|_{GUT}$ and
down-type Higgs mass $MH_d|_{GUT}$ which have been shown in  the conference
will be found in a shortly upcoming paper \cite{jbyannman}.

\begin{figure}[t]
\begin{center}
\begin{tabular}{cc}
\psfrag{ Universal mSugra}[c][c]{}
\includegraphics[width=0.5\textwidth]{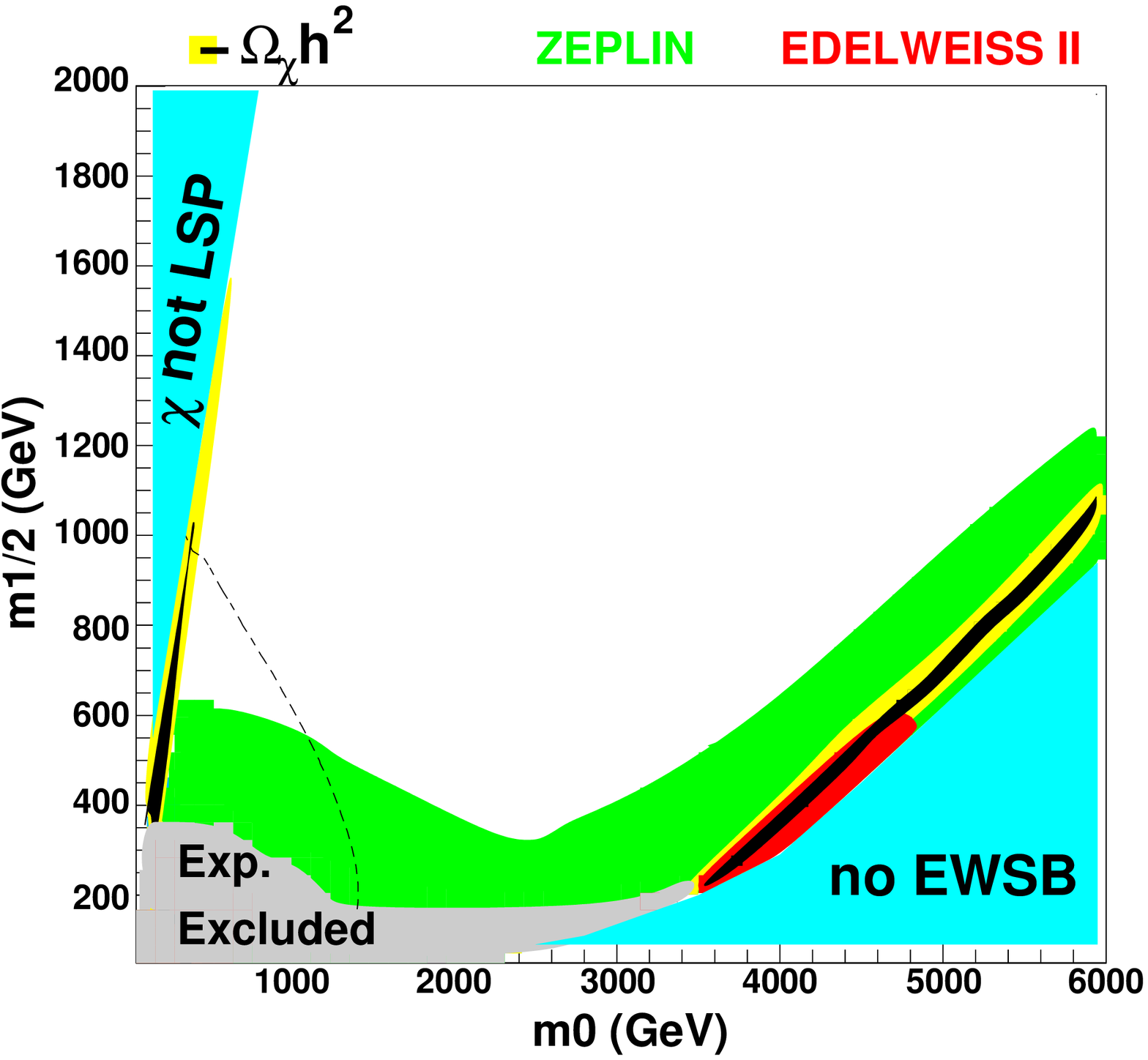}&
\psfrag{ Universal mSugra}[c][c]{}
\includegraphics[width=0.5\textwidth]{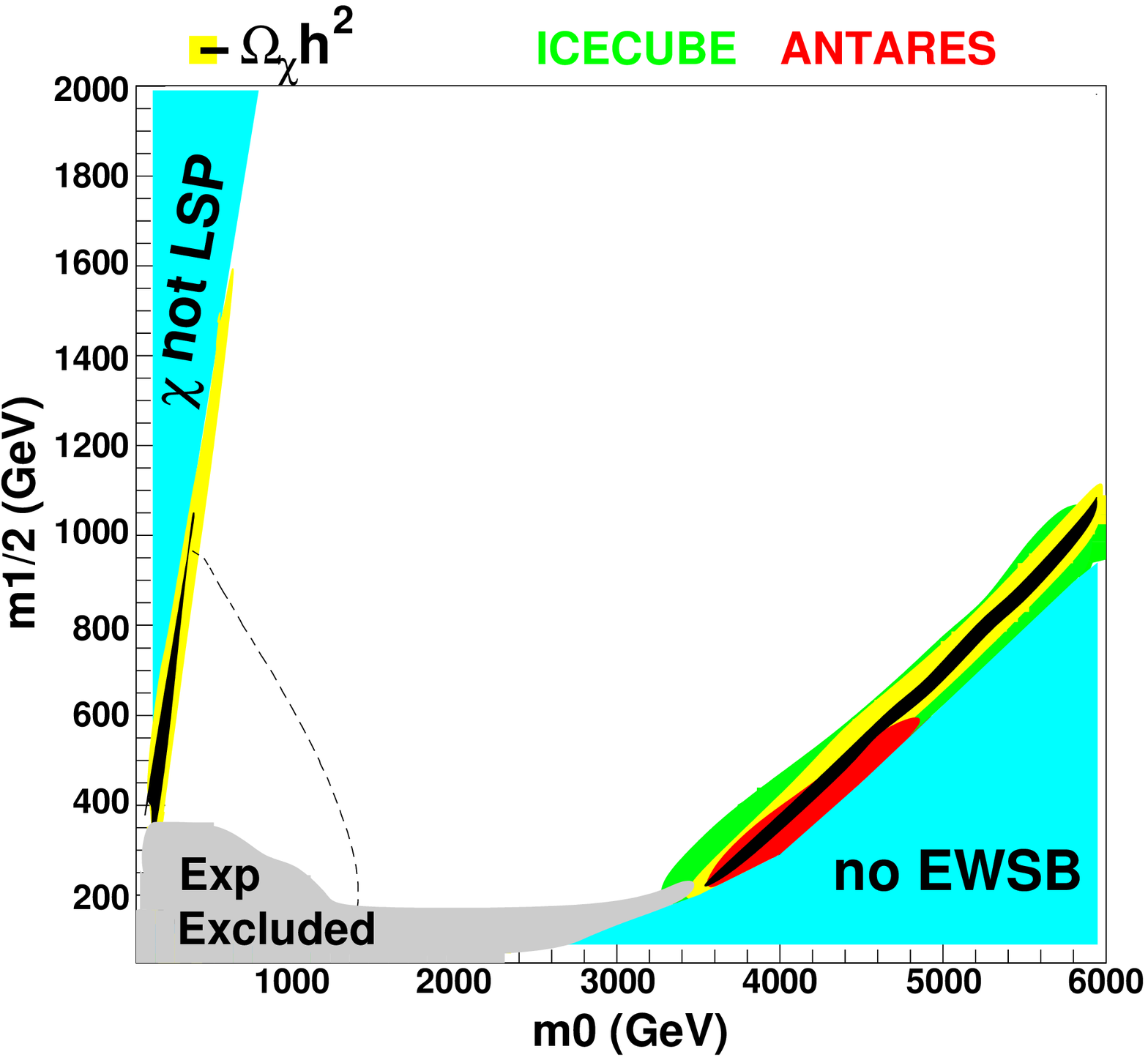}\\
a) Direct Detection & b) $\nu$ Indirect Detection (Sun) \\
\psfrag{ Universal mSugra}[c][c]{}
\includegraphics[width=0.5\textwidth]{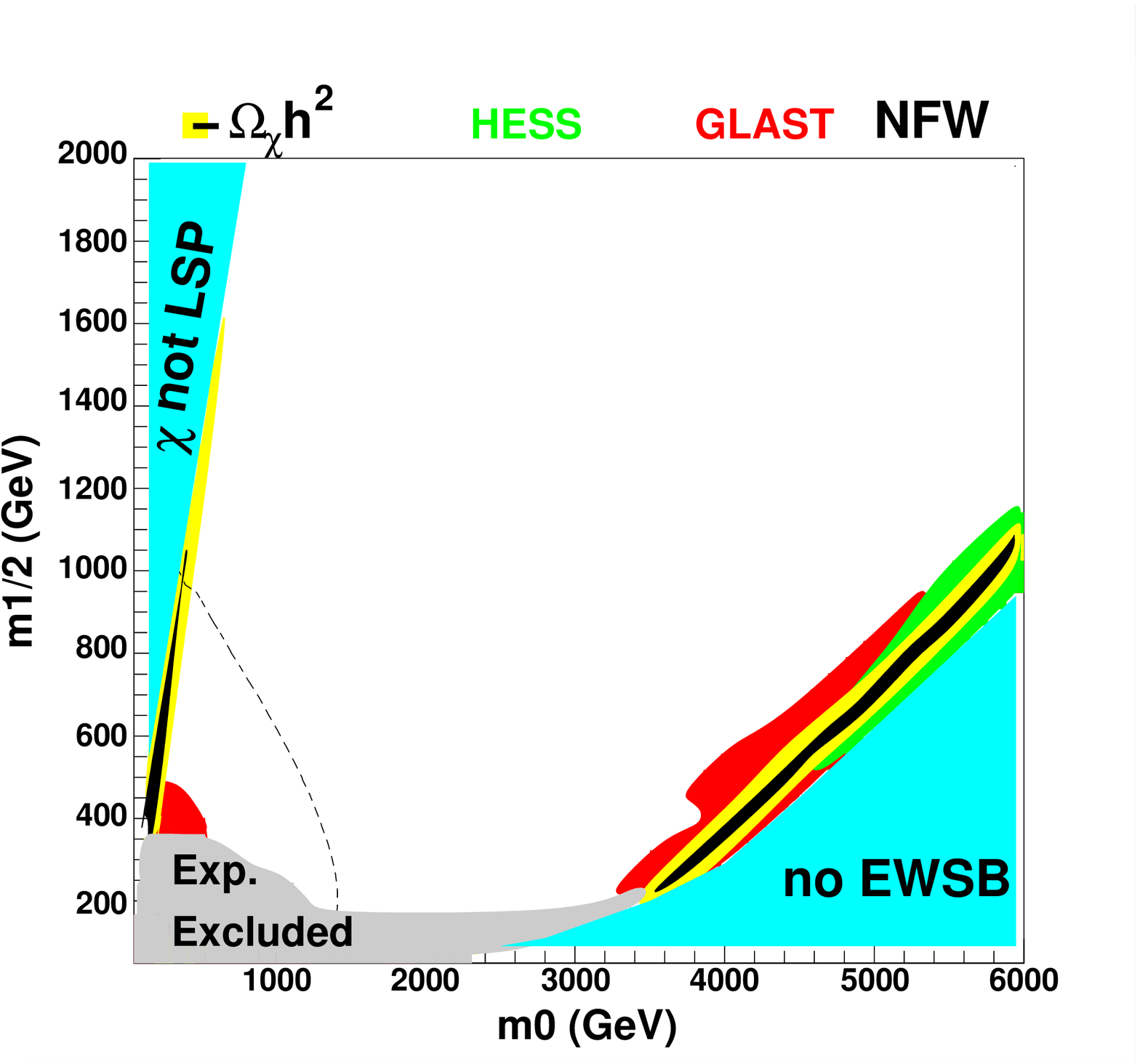}&
\psfrag{ Universal mSugra}[c][c]{}
\includegraphics[width=0.5\textwidth]{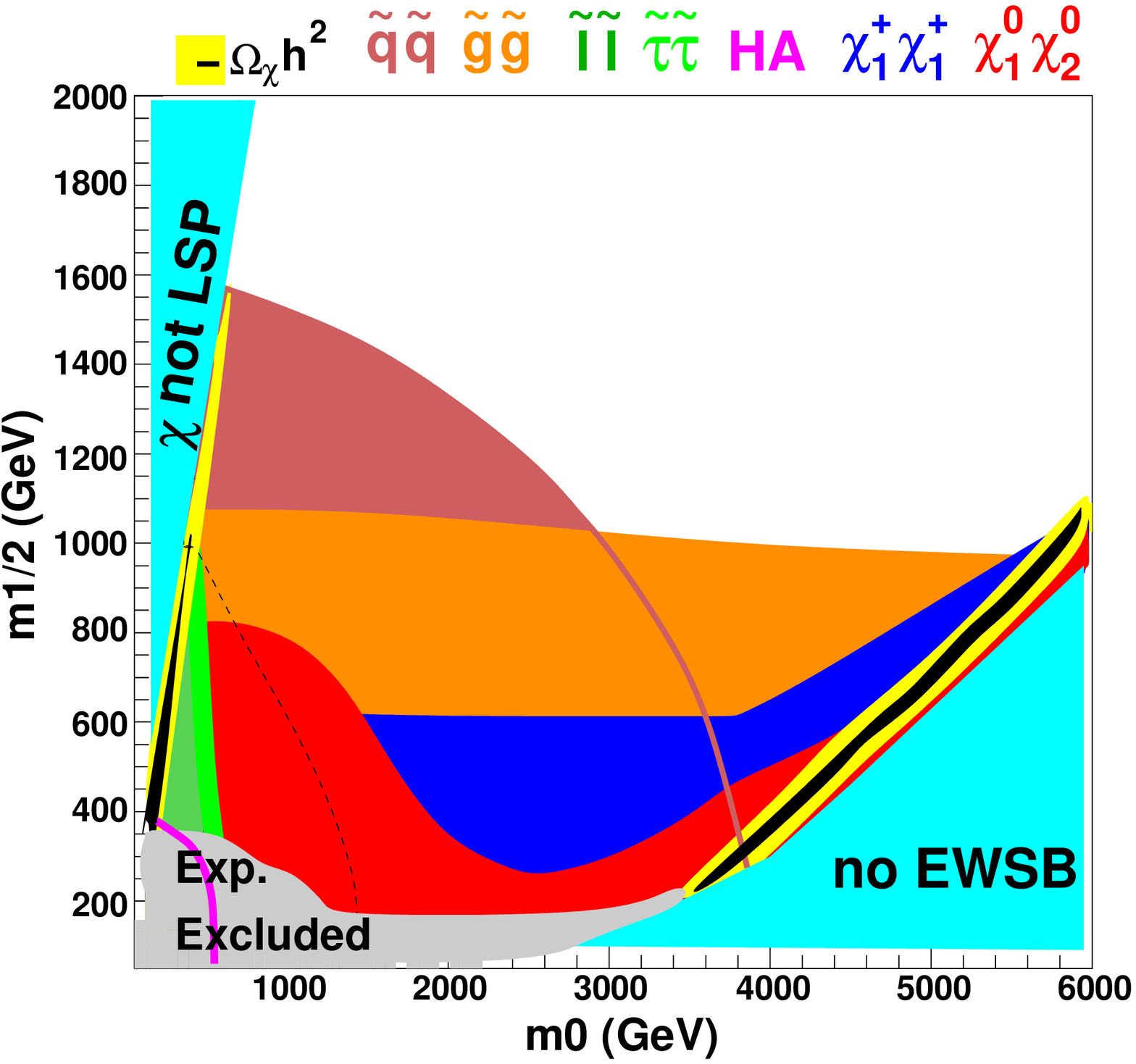}\\
c) $\gamma$ Indirect Detection (Galactic Center) & d) Collider production (LHC,LC)\\
\psfrag{ Universal mSugra}[c][c]{}
\includegraphics[width=0.5\textwidth]{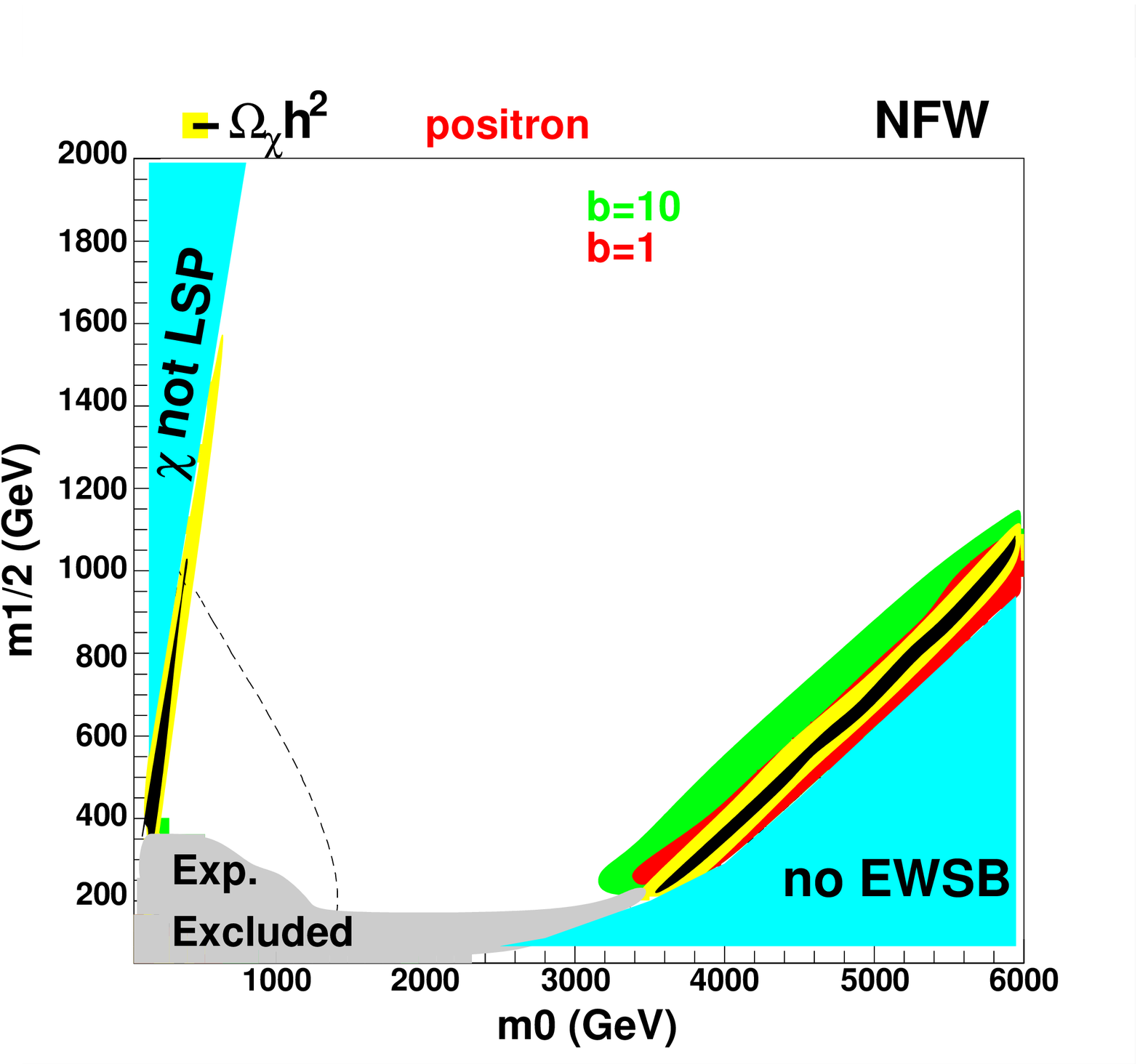}&
\psfrag{ Universal mSugra}[c][c]{}
\includegraphics[width=0.5\textwidth]{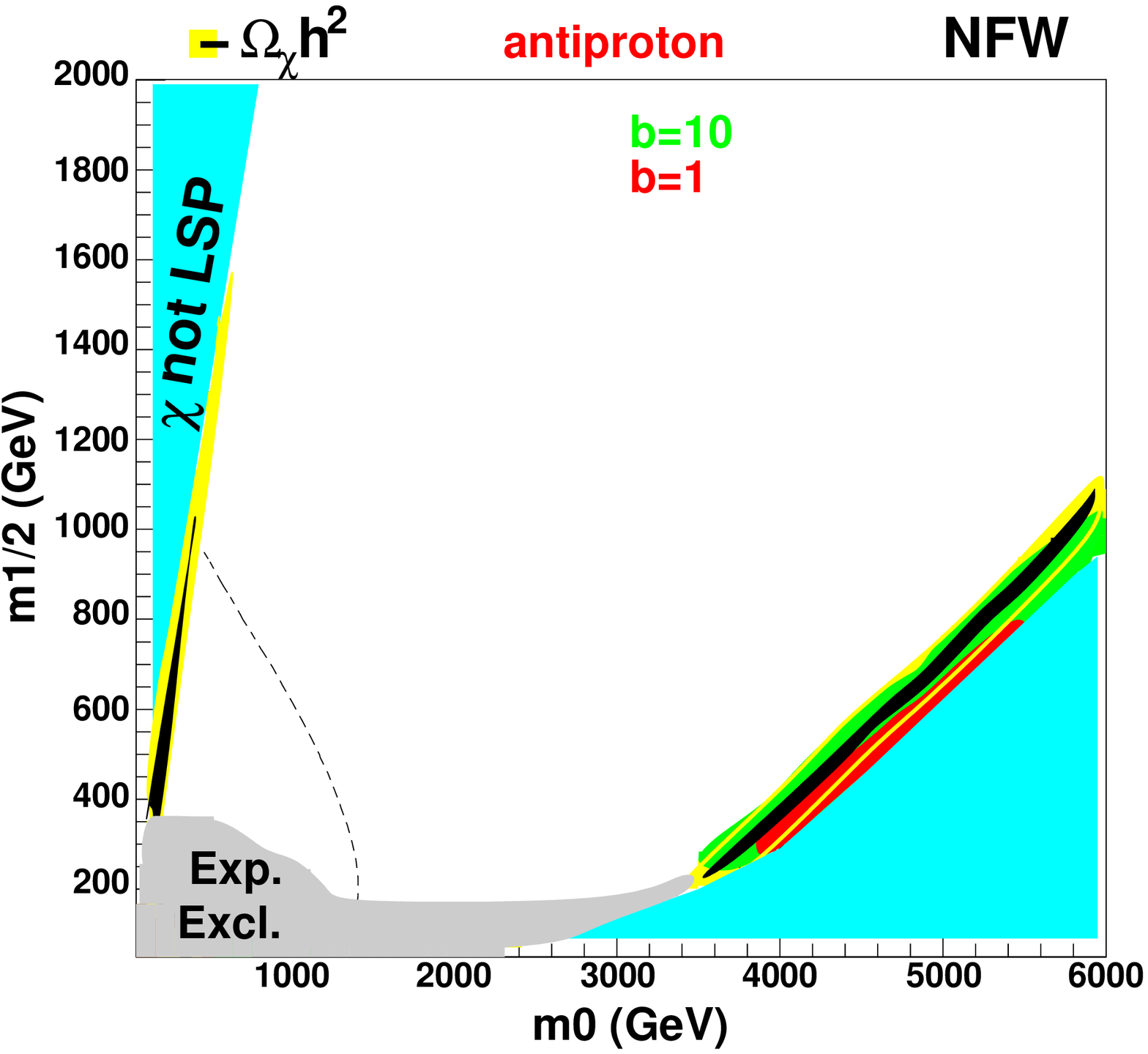}\\
e) $e^+$ Indirect Detection (halo) &f) $\bar{p}$ Indirect Detection (halo)\\
\end{tabular}
\caption{ MSUGRA Universal $A_0=0$, $\tan{\beta}=35$, $\mu>0$}
\label{fig:Utb35}
\end{center}
\end{figure}

\subsection{Universal case}

For intermediate values of $\tan{\beta}$, there are essentially 2 regions leading to interesting
neutralino relic abundance. The first one is near the stau LSP boundary (low $m_0$) thanks
to the $\tilde{\tau} \chi$ coannihilation processes. The second one is along
the boundary where the electroweak symmetry breaking cannot be achieved
radiatively ( Hyperbolic Branch/Focus Point : high $m_0$, low $\mu$) and where the neutralino is mixed bino-higgsino
enhancing $\chi \chi$ annihilation through $Z$ exchange and $\chi
\chi_1^+,\chi \chi_2^0$ coannihilations. Those two regions are generically thin.

Direct detection is then favored for light heavy Higgs scalar $H$ (mainly low $m_0,m_{1/2}$) or
 when the higgsino fraction is not negligible (high $m_0$) enhancing the
 neutralino-Higgs coupling (see Fig. \ref{fig:Utb35}a)). 

Concerning indirect detection with neutrino telescopes, a significant signal
from the Sun requires a relevant higgsino fraction to
enhance the capture rate through the spin dependent interaction $\chi
q\xrightarrow{Z} \chi q$ . This takes place only in the HB/FP branch (high
$m_0$, low $\mu$) where especially a ${\rm km^3}$ size detector is able to probe models satisfying the WMAP constraint (see
 Fig. \ref{fig:Utb35}b)).

Gamma indirect detection of neutralino annihilation in the Galactic Center
requires high annihilation cross section. The possible processes are either $\chi\chi
\xrightarrow{A}b\bar{b}$ which needs low $A$ (low $m_0,m_{1/2}$ )
and/or coupling ($\propto z_{11(2)}z_{13(4)}$) enhancement through higgsino
  fraction (high $m_0$) 
or $\chi\chi
\xrightarrow{Z}t\bar{t}$ annihilation which requires  an higgsino amount
as the coupling is $\propto z_{13(4)^2}$ (this is located near the EWSB
boundary{\it i.e} high $m_0$). This is shown on
 Fig. \ref{fig:Utb35}c) for a NFW halo profile ($\rho(r)\propto 1/r$ at small $r$).

The positron and antiproton fluxes have
essentially the same particle physics dependence ($\langle \sigma v \rangle$)
as gamma fluxes. The
favored regions for positron and antiproton are also where the neutralino
annihilation is strong (see
 Figs. \ref{fig:Utb35}e) and f) ). 

Collider production situation is shown on Fig. \ref{fig:Utb35}d). LHC is favored for low $\tilde{q}$ masses
( "low" $m_0$ values $\lesssim 2-2.5 $ TeV) and/or light $\tilde{g}$  (small
$M_3|_{low energy}$ {\it i.e}  $m_{1/2}\lesssim 1000$). 

Sleptons productions at the Linear Collider can be probe for low $\tilde{l}$
masses ($m_0\lesssim 700$ GeV, $m_0\lesssim 1000$ GeV). The $\chi \chi_2^0$
 (mainly bino and wino respectively) production is also favored for low $m_0$ through selectron exchange but
decreases when $m_{\tilde{e}}$ (mainly $m_0$) increase up to $m_0\sim 2000$
GeV where the higgsino fraction of the neutralinos allows the $Z$ exchange 
along the EWSB boundary. The chargino production follows first the kinematic
limit of wino chargino production ($m_{1/2}\sim 600$ GeV,
$2*m_{\chi^+_1}\simeq 2*0.8*m_{1/2}\simeq 1$ TeV) and then reaches higher
$m_{1/2}$ values thanks to the higgsino component of $\chi^+_1$ along the EWSB
boundary at high $m_0$. The region favorable to $HA$ production is restricted
to the low-left corner of the plane as $m_{A(H)}$ increase with both $m_0$ and $m_{1/2}$.

\subsection{The gluino mass : $M_3|_{GUT}$}

The gluino mass parameter has the main influence on MSSM spectrum through
$\alpha_s$ in RGE's \cite{Mynonuniv}. It decreases squark masses, increases the
up-type Higgs mass $M^2_{Hu}$  at low energy (less negative) and  decreases the
down-type Higgs mass $M^2_{Hd}$  which implies 
lighter $m_{A,H}$ and an increasing
of higgsino content of neutralinos and charginos as can be understood by tree
level relations: $
\mu^2\simeq-M^2_{H_u}-1/2M^2_Z \ \ {\rm and} \ \ m^2_A\simeq
M^2_{H_d}-M^2_{H_u}-M^2_Z$.

\begin{figure}[t]
\begin{center}
\begin{tabular}{cc}
\psfrag{non Universal : M3=0.6m1/2}[c][c]{}
\includegraphics[width=0.5\textwidth]{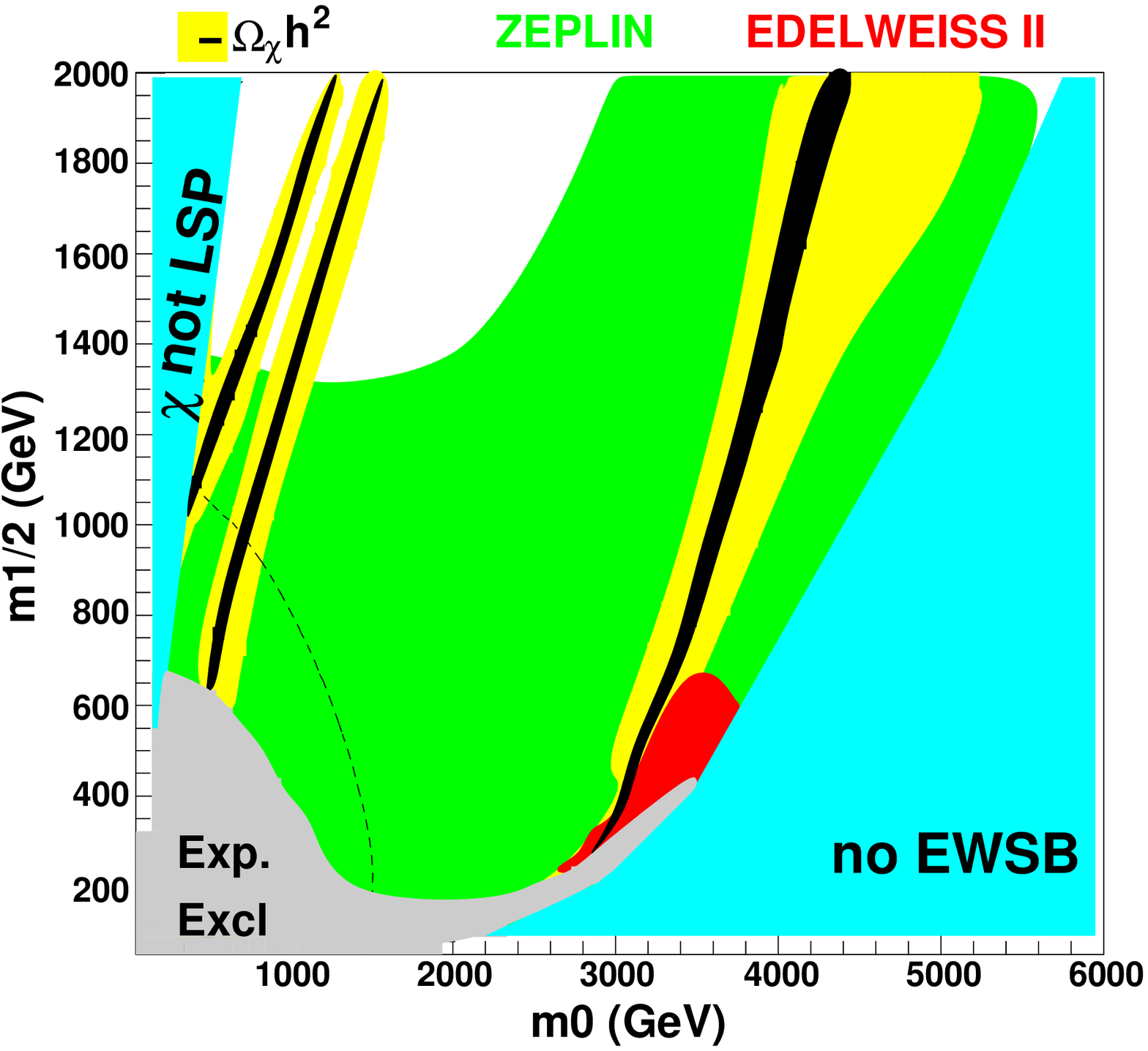}&
\psfrag{ non Universal : M3=0.6m1/2}[c][c]{}
\includegraphics[width=0.5\textwidth]{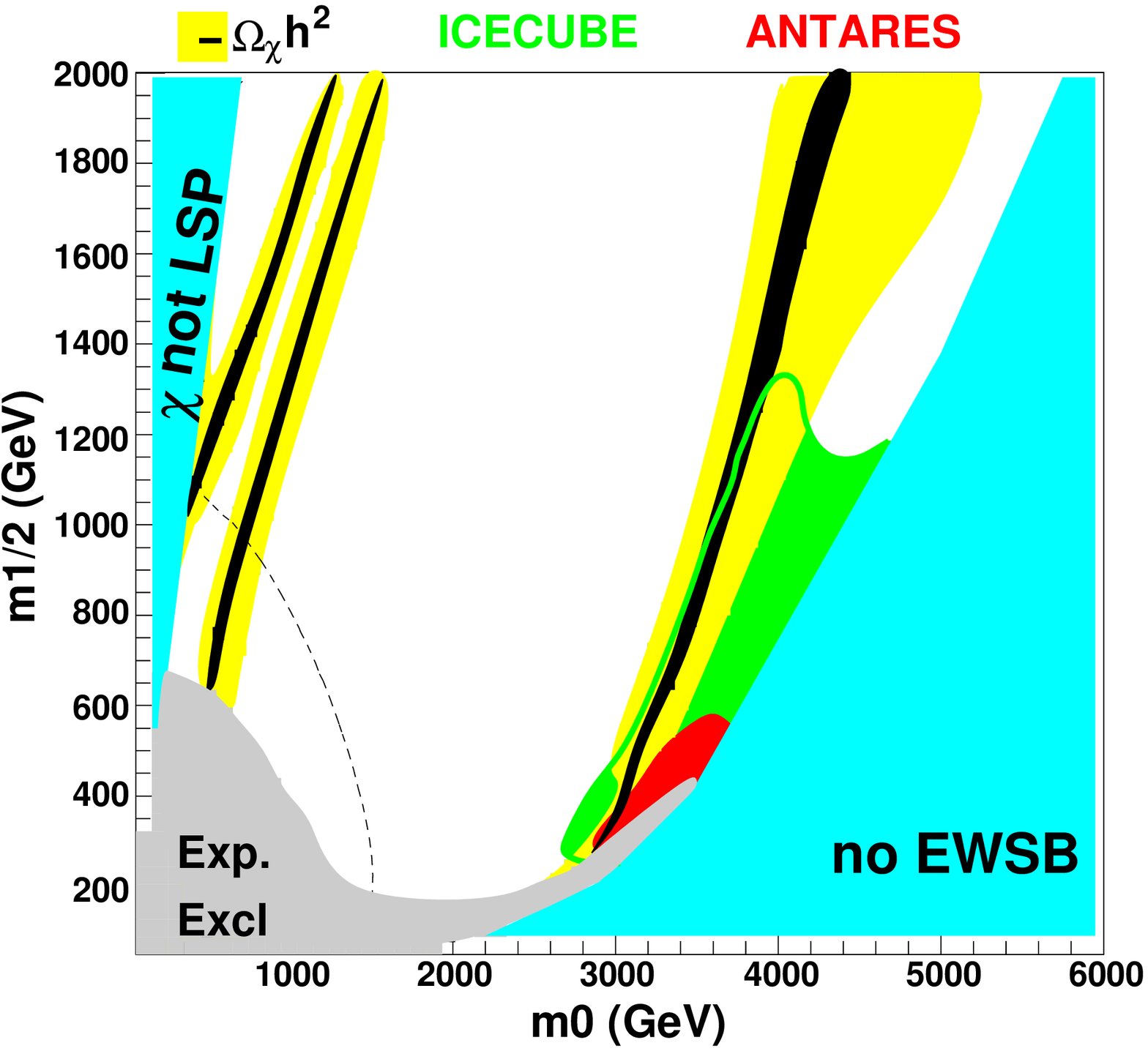}\\
a) Direct Detection & b) $\nu$ Indirect Detection (Sun) \\
\psfrag{ non Universal : M3=0.6m1/2}[c][c]{}
\includegraphics[width=0.5\textwidth]{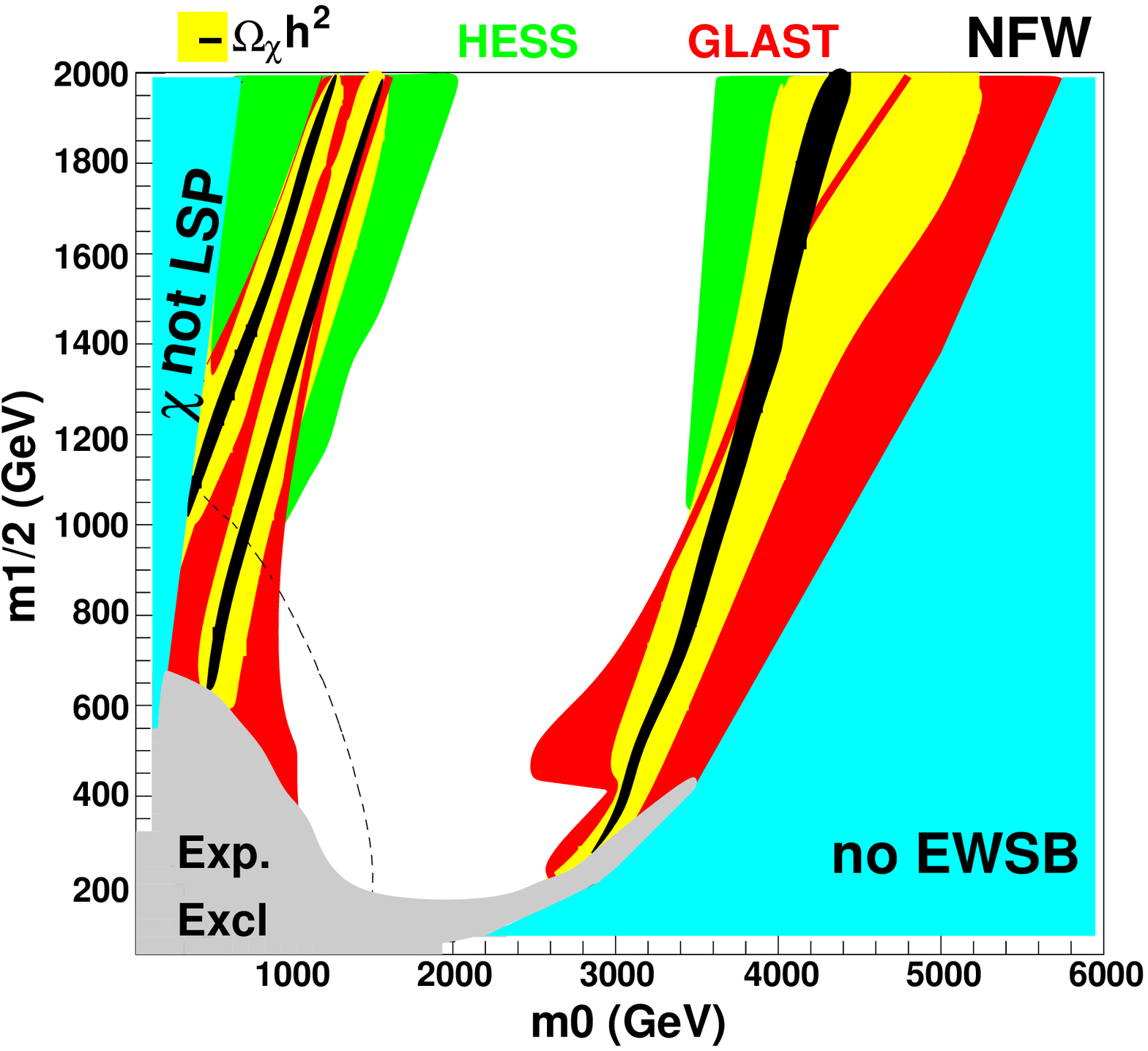}&
\psfrag{ non Universal : M3=0.6m1/2}[c][c]{}
\includegraphics[width=0.5\textwidth]{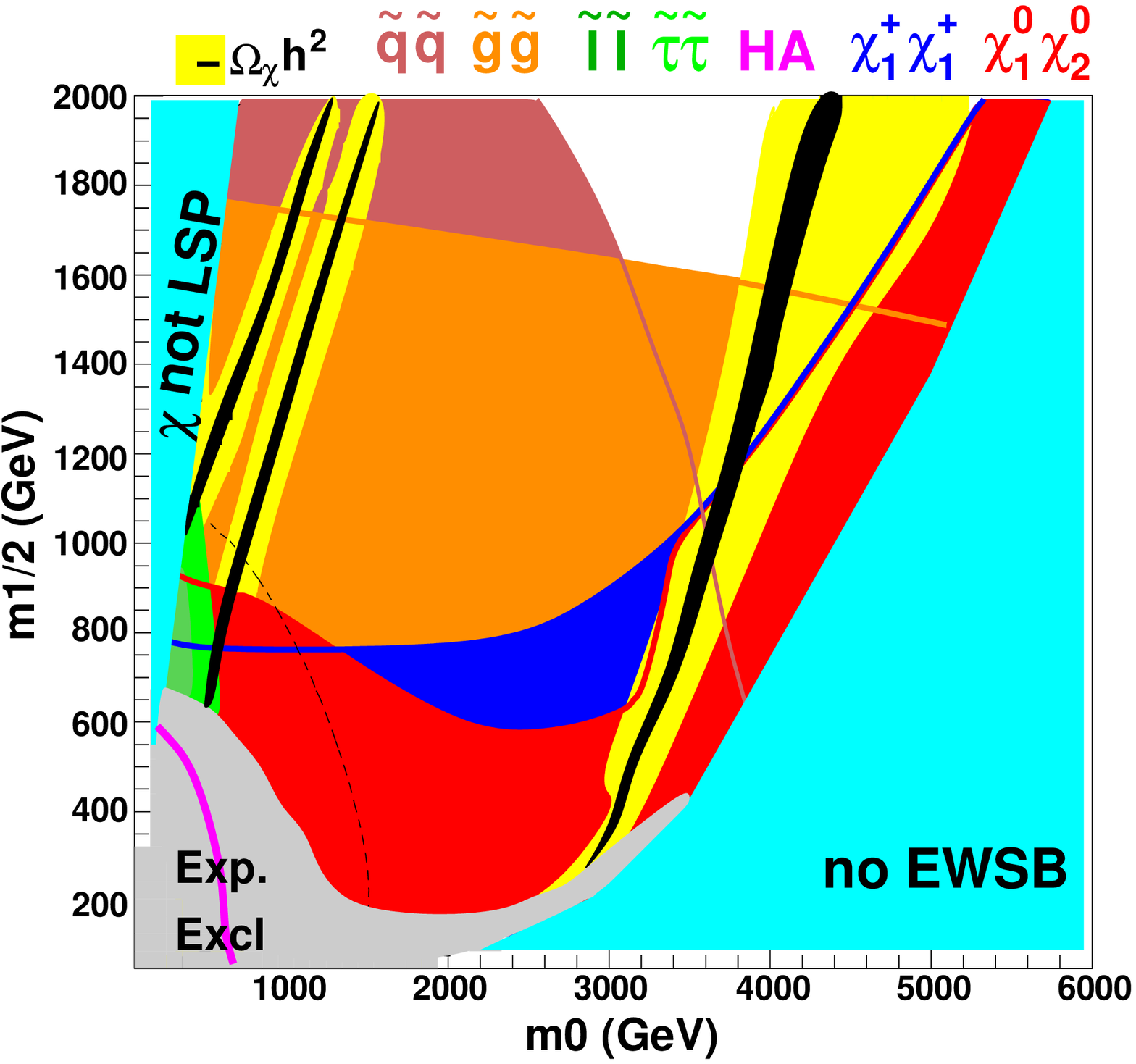}\\
c) $\gamma$ Indirect Detection (Galactic Center) & d) Collider production (LHC,LC)\\
\psfrag{ non Universal : M3=0.6m1/2}[c][c]{}
\includegraphics[width=0.5\textwidth]{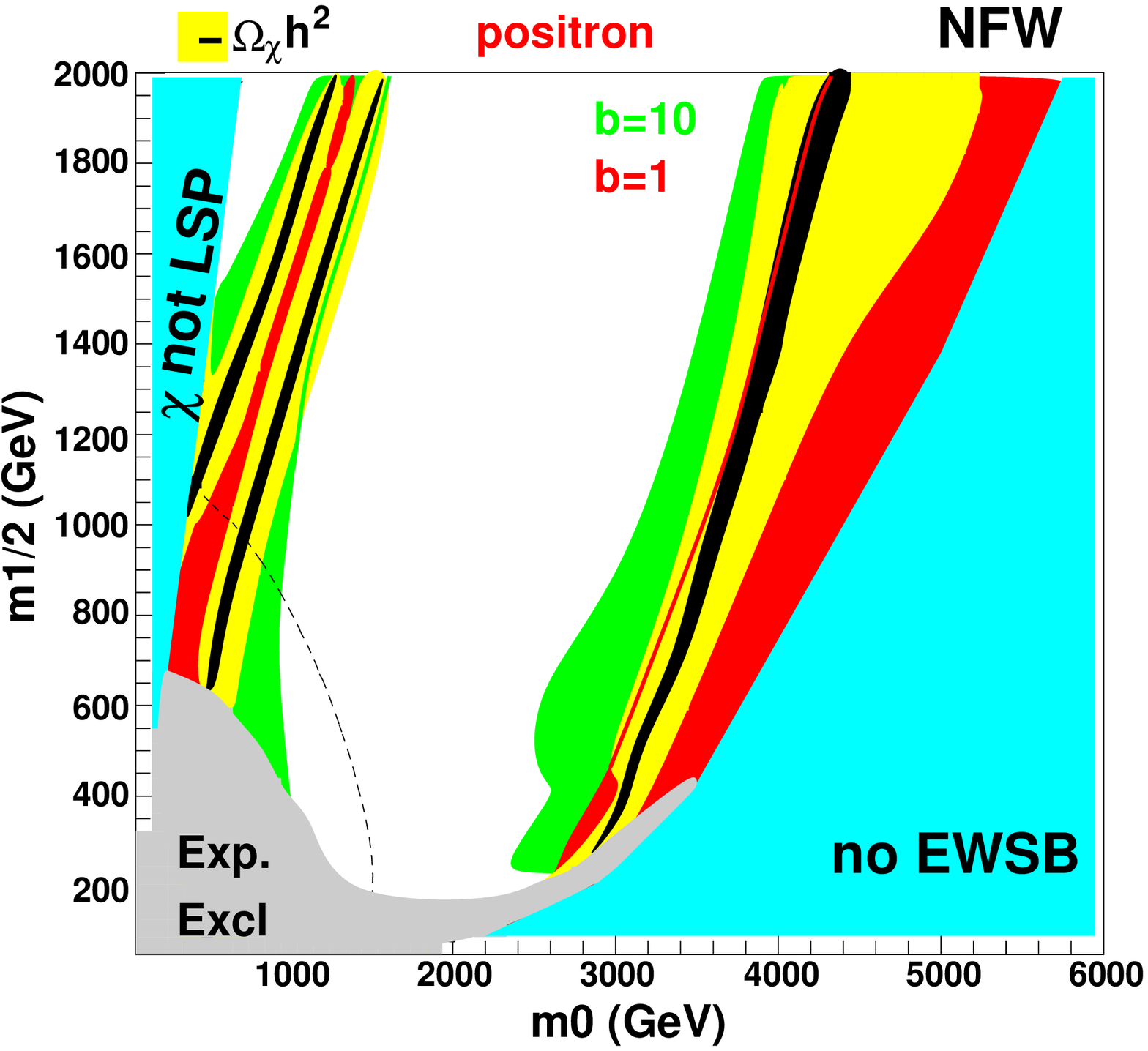}&
\psfrag{ non Universal : M3=0.6m1/2}[c][c]{}
\includegraphics[width=0.5\textwidth]{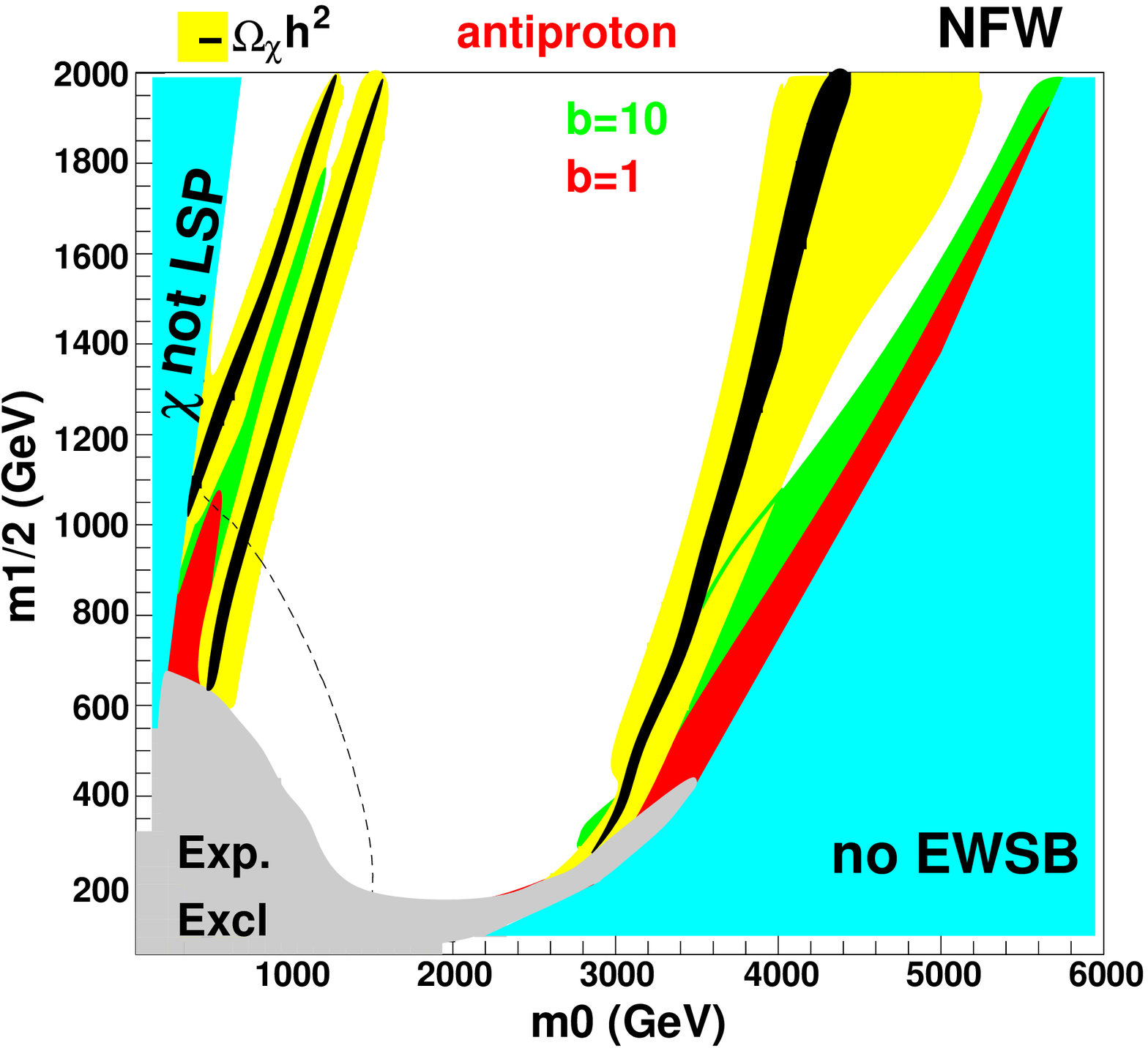}\\
e) $e^+$ Indirect Detection (halo) &f) $\bar{p}$ Indirect Detection (halo)\\
\end{tabular}
\caption{ MSUGRA non Universal gluino mass $M_3|_{GUT}=0.6m_{1/2}$, $A_0=0$, $\tan{\beta}=35$, $\mu>0$}
\label{fig:M3tb35}
\end{center}
\end{figure}

As a result, relic density constraints are more easily satisfied than in the
universal case: both $\chi \chi
\xrightarrow{A}b\bar{b}$ annihilation (thanks to higher coupling {\it and}
lighter $A$ which can open the $A$ funnel) and focus point region with $\chi \chi
\xrightarrow{Z}t\bar{t}$ annihilation are enhanced.
Direct detection get advantage of better couplings $z_{11}z_{13}$ and lighter
$H$. The higher higgsino fraction favors neutrino indirect detection in the
coupling in $\chi q \xrightarrow{Z} \chi q$ of the capture rate.
Gamma,positron and antiproton indirect detection are favored by the
annihilation  enhancement. The $\chi \chi \xrightarrow{A}
b\bar{b}$ process is favored by the higgsino fraction and the resulting
better couplings $z_{11}z_{13}$ as well as a lighter pseudoscalar $A$. The  $\chi \chi \xrightarrow{Z}
t\bar{t}$  
process couplings, $\propto z_{13(4)}^2$, are favored by the higgsino fraction as
well as $\chi \chi \xrightarrow{\chi^+}
W^+ W^-$.
LHC gets strong potentiality enhancement thanks to lighter squarks
($\tilde{t}$) and lighter
gluinos.
For the Linear Collider, the $HA$ production is favored (lighter $H,A$) as
well as  $\chi^+\chi^-$ $\chi\chi^0_2$ favored by the lower $\mu$ values.
The non universal $M_3|_{GUT}$ situation is illustrated on
Fig. \ref{fig:M3tb35} (compare panels with those of
Fig. \ref{fig:Utb35}).

This kind of models with light gluino mass at GUT scale are very favorable for SUSY
detection in colliders as well as all neutralino dark matter searches and can be found in
some effective string inspired scenarios \cite{PBmodels}.

\section{Conclusion}

Dark matter experiments and collider searches will be a quite conclusive step 
to probe the possibility of low energy supersymmetry and neutralino dark
matter in the Minimal Supersymmetric Standard Model. The possible 
correlation  between (non) signals of different kinds of detection will bring
the maximum of information on models and scenarios both for supersymmetry and 
astrophysics.

\section*{Acknowledgments}{I'm grateful to the organizers for this conference.
  I thank my collaborators on this work J.-B. De Vivie and
  Y. Mambrini. This work is supported by the I.I.S.N. and the 
Belgian Federal Science Policy (return grant and IAP 5/27)}

%==============================================================================

\section*{References}

\nocite{}
\bibliography{bmn}
\bibliographystyle{unsrt}

\end{document}